\documentclass[aps,prl,showpacs,noshowkeys,amsmath,amssymb,amsfonts,reprint]{revtex4-1}
\usepackage{graphicx}
\usepackage{dcolumn}
\usepackage{bm}
\begin{document}
\title{Defect Dynamics in Artificial Colloidal Ice:  Real-Time Observation,\\ Manipulation and Logic Gate}
\author{Johannes Loehr$^{1,2}$, Antonio Ortiz-Ambriz$^{1,3}$ and Pietro Tierno$^{1,3}$}
\email{ptierno@ub.edu}
\affiliation{
$^1$Departament de F\'isica de la Mat\`eria Condensada, Universitat de Barcelona, Barcelona, Spain\\
$^2$Physikalisches Institut, Universit\"at Bayreuth, 95440 Bayreuth, Germany\\
$^3$Institut de Nanoci\`encia i Nanotecnologia, IN$^2$UB, Universitat de Barcelona, Barcelona, Spain
}
\date{\today}
\begin{abstract}
We study the defect dynamics in a colloidal spin ice system realized by filling a square lattice of topographic
double well islands with repulsively interacting magnetic colloids.
We focus on the contraction of defects in the ground state, and contraction/expansion
in a metastable biased state.  
Combining real-time experiments with simulations,
we prove that these defects behave like emergent topological monopoles
obeying a Coulomb law with an additional
line tension.
We further show how to realize a completely resettable "NOR" gate, which provides guidelines 
for fabrication of nanoscale logic devices based on the motion of topological magnetic monopoles.
\end{abstract}
\maketitle
Geometric frustration is a complex phenomenon which encompasses 
a broad range of systems, from magnetic materials~\cite{Ste01}, 
to ferroelectrics~\cite{Cho11}, trapped ions~\cite{Ki10}, 
confined microgel particles~\cite{Han08}, and folding proteins~\cite{Bry87}.
It emerges when the spatial arrangement of the system elements cannot 
simultaneously minimize all interaction energies, and leads to 
exotic phases of matter with a low-temperature degenerate ground
state, such as spin ice~\cite{Har97,Ram99,Moe06}.
Artificial spin ice systems (ASI) are lattices of interacting 
nanoscale ferromagnetic islands, recently introduced as a versatile 
model to investigate geometrically frustrated states~\cite{Wan06,Nis13}, 
including the role of disorder~\cite{Dau11,Bud12}, thermalization~\cite{Arn12,She13,Far13}
and the excitation dynamics~\cite{Lad10,Mor11,Men11,Pha11,Kap14}.
In opposition to bulk spin ice such as pyrochlore compounds, ASI enables to 
directly visualize the spin textures and to tailor the spatial arrangement 
of the system elements.\\
An intriguing aspect in ASI, which is attracting much theoretical interest, is the dynamics of 
defects~\cite{Mol09,Mol10,Sil12,Nas12,Sil13,Lev13,Ved16,Bha16}.
The interactions between pairs of defects is one of the distinctive
features between three dimensional (3D) and two dimensional (2D) spin ice.
In a 3D pyrochlore compound, the spins are located 
on a lattice of corner-sharing tetrahedra, and can point either towards 
the tetrahedra center (spin {\it in}), or away from it (spin {\it out}).
Thus the ground state (GS) follows the ``ice-rules'', with two spins coming {\it in} and two going {\it out}
of each vertex in order to decrease the vertex energy.
At finite temperature, defects that behave like ``magnetic monopoles''~\cite{Bra11,Jau11}
can emerge when a spin flips, producing a local increase of the magnetic energy.
A way to overcome the system complexity is to use the ``dumbbell'' model~\cite{Cas08}, 
which only considers the magnetic charge distribution at the vertices of the lattice.
Within this formalism, it was shown that in 3D spin ice, a pair 
of defects connected by strings of flipped spins only interact 
through a magnetic Coulomb law at low temperature.
In contrast, numerical simulations show that 
for a 2D square ASI, {\it i.e.} a projection of the 3D 
ice system on a plane, such a string requires an additional energetic 
term in form of a line tension~\cite{Mol09}.
The reason is that, while in a 3D system all
spin configurations that satisfy the ice rules have equal energy, 
in the 2D square ASI the distance at a vertex between opposing spins 
is greater than the distance between adjacent spins.
This results in a lift of the degeneracy of the ground state, which is 
now represented by a two-fold degenerate antiferromagnetic order.\\
String tension and the Coulombic interactions in ASI have
been calculated by Monte Carlo simulations~\cite{Mol10,Sil12,Ved16}, 
however direct experimental measurements remain elusive.
The difficulty of preparing the system in the GS and the extremely fast spin dynamics in nanoscale ASI makes real-time observation
challenging, suggesting the use of alternative systems.
Here we overcome these limitations by realizing an artificial 
colloidal spin ice system,
a microscale soft matter analog of a frustrated nanoscale ASI.
In this system we investigate the 
real-time dynamics of monopole-like defects via experiments and numerical simulations, and directly measure the line tension
and Coulombic contributions.
Further we demonstrate defect manipulation via external field, and realize a logic operation based on
magnetic current.\\
Our experimental system is inspired by previous theoretical works on 
electrostatically interacting colloids in bistable optical traps~\cite{Lib06,Lib12}.
The schematic in Fig.1(a) and the experimental realization 
in Fig.1(b) illustrate the main idea.
By soft lithography, we realize a square lattice of bistable topographic traps with lattice constant $a=29 \, {\rm \mu m}$.
Each trap is composed of two wells of depth $\sim 3 {\rm \mu m}$, connected by a small hill at the middle
with elevation $\langle h \rangle = 0.86 {\rm \mu m}$, Fig.1(d-f)~\cite{EPAPS}.
These traps are designed to confine a colloidal particle in one of the two sides, such that the particle can
cross the hill when subjected to an external force, but it cannot escape from the bistable confinement.
We induce repulsive interactions by using paramagnetic colloids
with diameter $d=10.3 \, {\rm \mu m}$ and magnetic volume 
susceptibility $\chi=0.08$ (Microparticles GmbH).
Under an external magnetic field perpendicular to the particle plane, $\bm{B}=B_z \hat{\bm{z}}$, 
the colloids repel by a tunable pair potential, $U^m_{ij}=\omega       \frac{a^2}{r^3_{ij}}$, where
$\omega=\mu_0 m^2/(4\pi a^3)$ is the coupling constant
with $\bm{m}=\pi d^3 \chi \bm{B}/(6\mu_0)$ the induced moment, $r_{ij}=|\bm{r}_i-\bm{r}_j|$, and $r_i$
is the position of particle $i$.
The gravitational potential for a particle to jump a
hill is $U^{hill}_g=910 k_B T$, and $U^{wall}_g=3740 k_B T$ 
to leave the bistable trap. Here $k_B$ is the Boltzmann 
constant, $T=293 {\rm K}$ and we apply the external 
field such that $U^{hill}_g<U^{m}<U^{out}_g$~
\footnote{To calculate the gravitational potentials we 
use the density mismatch $\Delta \rho =0.9 \, {\rm g cm^{-3}}$
between the particle and the suspending medium. Optical tweezers are used to load one particle per double well, to order the system or create defects.}.\\
Once filled with one particle per double well, one can assign a vector
(analogous to a spin) to each particle, such that it points
from the free well to the well occupied by the particle.
As shown in Fig.1(c), it is possible to construct a set of
ice rules for the colloidal artificial ice similar to the nanoscale ASI~\cite{Lib06,Nis14}.
Vertices with three ($S_{V}$) or four ($S_{VI}$) colloids {\it in} are energetically unfavourable, and
they are topologically connected with low energy vertices having three ($S_{II}$) or four ($S_{I}$) colloids {\it out}.
\begin{figure}[t]
\begin{center}
\includegraphics[width=\columnwidth,keepaspectratio]{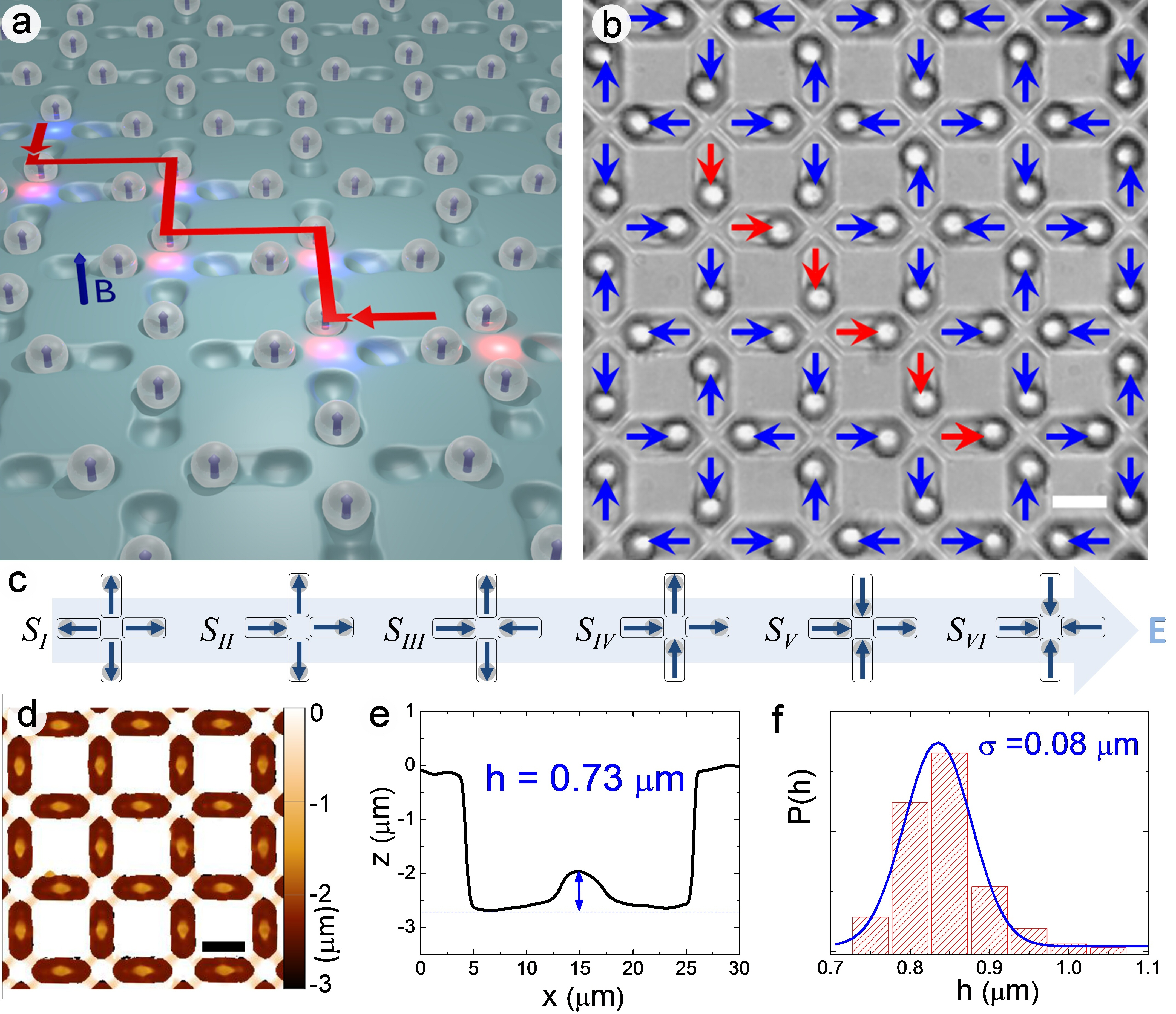}
\caption{(color online)
(a) Schematic showing the colloidal spin ice composed by interacting colloids
in a square lattice of double-wells.
The gray (red) line shows a defect line separating two $q  = \pm 2$ defects in the GS.
(b) Microscope image of an experimental defect line in a square lattice of lithographic double wells
filled with paramagnetic colloids.
Black (blue) arrows denote spin directions, gray (red) arrows highlight the defect line. 
Scale bar is $15 {\rm \mu m}$
(c) Vertex configurations for the colloidal square ice. Vertex energy increases from left to right.
(d) Optical profilometer image of the lithographic square lattice.
(e) Cross-section of a typical double well characterized by a central hill of height $h=0.73 \mu m$.
(f) Distribution of hill height $h$ fitted with a Gaussian function (continuous line).
}
\label{figure1}
\end{center}
\end{figure}
Thus the GS is composed by $S_{III}$ vertices~\cite{Ort16},
while the metastable biased state 
has high energy $S_{IV}$ vertices.
Both configurations satisfy the ice rules.
According to the ``dumbbell'' model~\cite{Cas08}, we can associate 
to each spin a "magnetic charge", which is positive (negative) for spin {\it in} ({\it out}).
The total charge at each vertex $i$ is given by the sum over 
all neighbouring spins $q=\sum_i q_i$, and both the GS and the biased state correspond to $q=0$,
while all other vertices have a net charge.\\
We start by analyzing the contraction of a pair of $q=\pm 2$ ($S_{II}$ and $S_V$) charged defects connected by
a line of six flipped spins along the diagonal in the GS, Fig.1(b) and 2(a)~\footnote{The small
system size was chosen in order to both
make more evident the effect of Coulumbic-like interactions,
and to minimize the breaking of the defect line caused by
small disorder related with the distribution of hills, Fig.1(f).}.
After preparing the system with the optical tweezers, we 
switch the field on and measure the relaxation toward equilibrium.
As shown in Fig.2(a) and VideoS1 in~\cite{EPAPS}, both defects approach via
a stepwise flipping of the colloids position and the system recovers the GS.
Theoretical work~\cite{Mol10} based on the dumbbell model~\cite{Cas08} predicts
the interaction potential between the two defects in the 2D ASI as, $V(l)=-Q/l+\kappa l+c$.
Here $Q$ is the topological Coulombic charge, $\kappa$ the
line tension, and $c$ a constant associated with the creation of the defect pairs~\cite{Sil13}.
We confirm the validity of this assumption in our system, by explicitly calculating the 
energy cost $V(l)=E_{exc}(l)-E_{GS}$ of a defect line of length $l$ in the GS, which 
can be obtained
by subtracting the GS energy from the energy of the excited configuration.
The magnetic energy is given by the sum of all dipole interactions as, $E= \sum_i\sum_{j\neq i} U^m_{ij}$.
In the inset of Fig.2(b), we show the normalized
potential $V(l)/\omega$. We subtract its linear part
in order to emphasize the presence of a magnetic Coulumbic term. 
Since $V(l)$ scales with the coupling constant, it follows that $Q, \kappa \sim\omega\sim H^2$.
\begin{figure}[t]
\begin{center}
\includegraphics[width=\columnwidth,keepaspectratio]{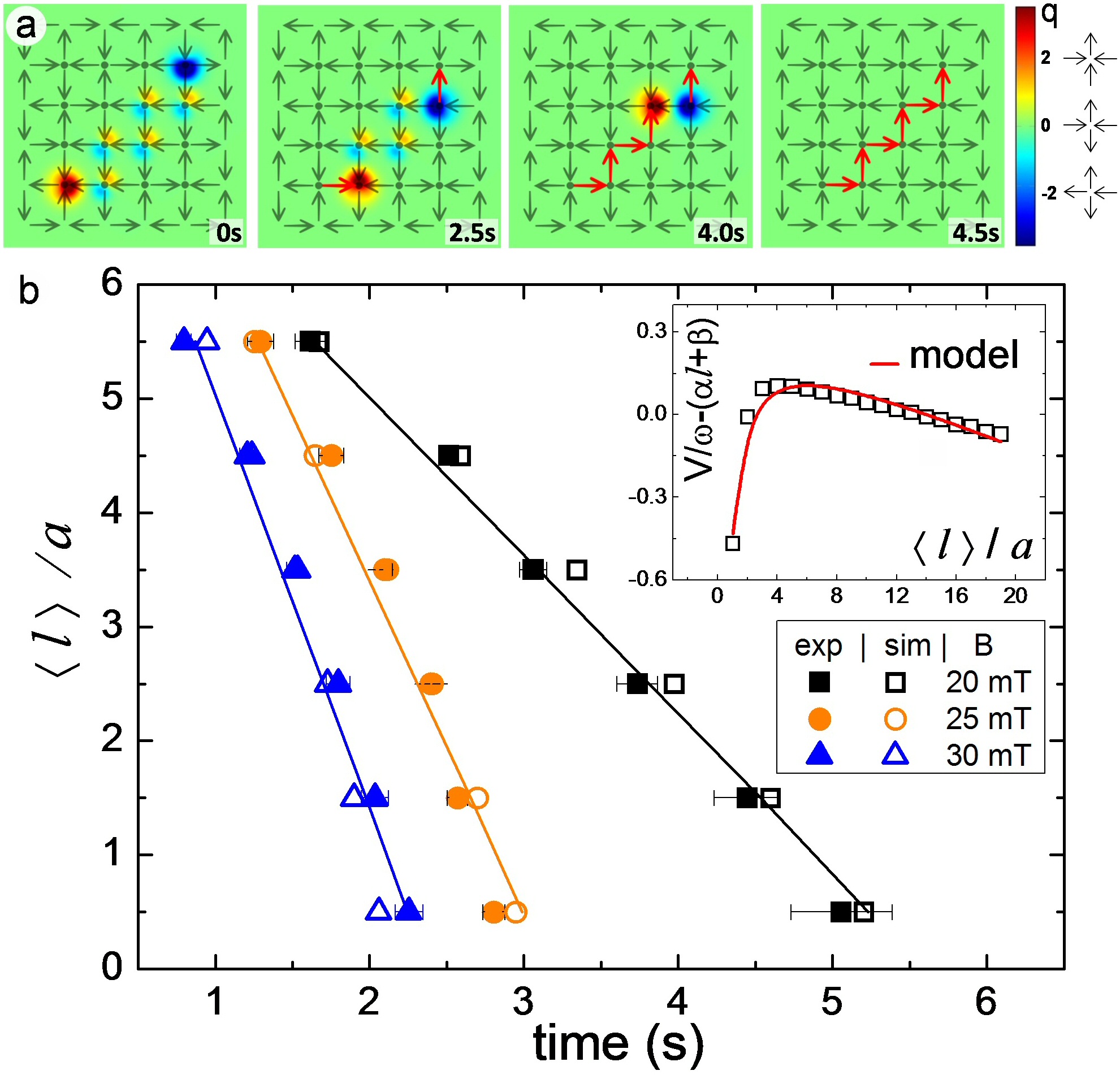}
\caption{(color online)
(a) Color map
showing the net vertex charges
in the experiments
for a defect line connecting two $q=\pm2$ defects
under a field $B_z=25.7 {\rm mT}$ (VideoS1 in~\cite{EPAPS}).
The line consists
of high energy $S_{IV}$ vertices with a zero charge
but a net dipole,
which give raise to the additional line-tension
term.
(b) Average line length $\langle l \rangle$
versus time
for three different magnetic fields.
Filled (empty) symbols denotes experiments (numerical simulation),
continuous lines are fit
from Eq.(1) in the text.
Inset:  normalized interaction potential $V(l)/\omega$
between two topological defects minus
its linear contribution ($\alpha l+\beta$).
Red line is a fit using the potential described in the text.
}
\label{figure2}
\end{center}
\end{figure}
By fitting this potential, we obtain the ratio 
$Q/\kappa=0.0290\pm 0.0014a^2$ between the Coulombic and line tension contribution,
which is one order of magnitude lower than the corresponding one found for ASI~\cite{Mol09}.\\
Figure 2(b) shows experiments and simulations of the average line length $\langle l \rangle$ 
obtained by measuring the particle residence time within the traps~\cite{EPAPS}.
We describe the dynamics of the defect line
with an overdamped equation of motion with a friction coefficient $\gamma$,
\begin{equation}
\gamma \frac{dl}{dt} = -\frac{\partial V}{\partial l} = -\frac{Q}{l^2}-\kappa \, \, .
\end{equation}
We assume negligible the thermal fluctuations given the large size of the employed particles,
and we justify our choice of overdamped dynamics, as opposed to the infradamped dynamics in nanoscale 
ASI~\cite{Ved16}, by checking that the defect motion effectively shows a velocity profile 
linear with the applied force~\cite{EPAPS}.
By solving Eq.(1), see~\cite{EPAPS}, we fit its solution to the experimental data in Fig.2(b).
We use the ratio $Q/\kappa$ obtained 
from the calculation of $V(r)$ (inset Fig.2(b)), and reduce to $Q/\gamma$ the sole unknown parameter.
Figure 2(b) shows the results of this procedure, confirming that the observed phenomena are well captured by Eq.(1).
In all our analysis  we use $\gamma$ as scaling factor for the topological Coulomb
charge, $Q$. However $Q$ may be estimated in first approximation by 
considering that the defects are composed by colloidal particles
approaching at a constant speed in a liquid medium. \footnote{The friction
between such particles is given by $\gamma=6\pi \eta d/2$, being
$\eta=10^{-3} {\rm Pa \cdot s}$ the viscosity of the medium (water).}
For an applied field of $B_z=30 {\rm mT}$, we obtain
for the colloidal spin ice $Q_M \sim \sqrt{4 \pi |Q|/\mu_0} =  5.7 \pm  1.5 \times 10^{-8} {\rm m/s}$.
To further validate our
\begin{figure}[b]
\begin{center}
\includegraphics[width=\columnwidth,keepaspectratio]{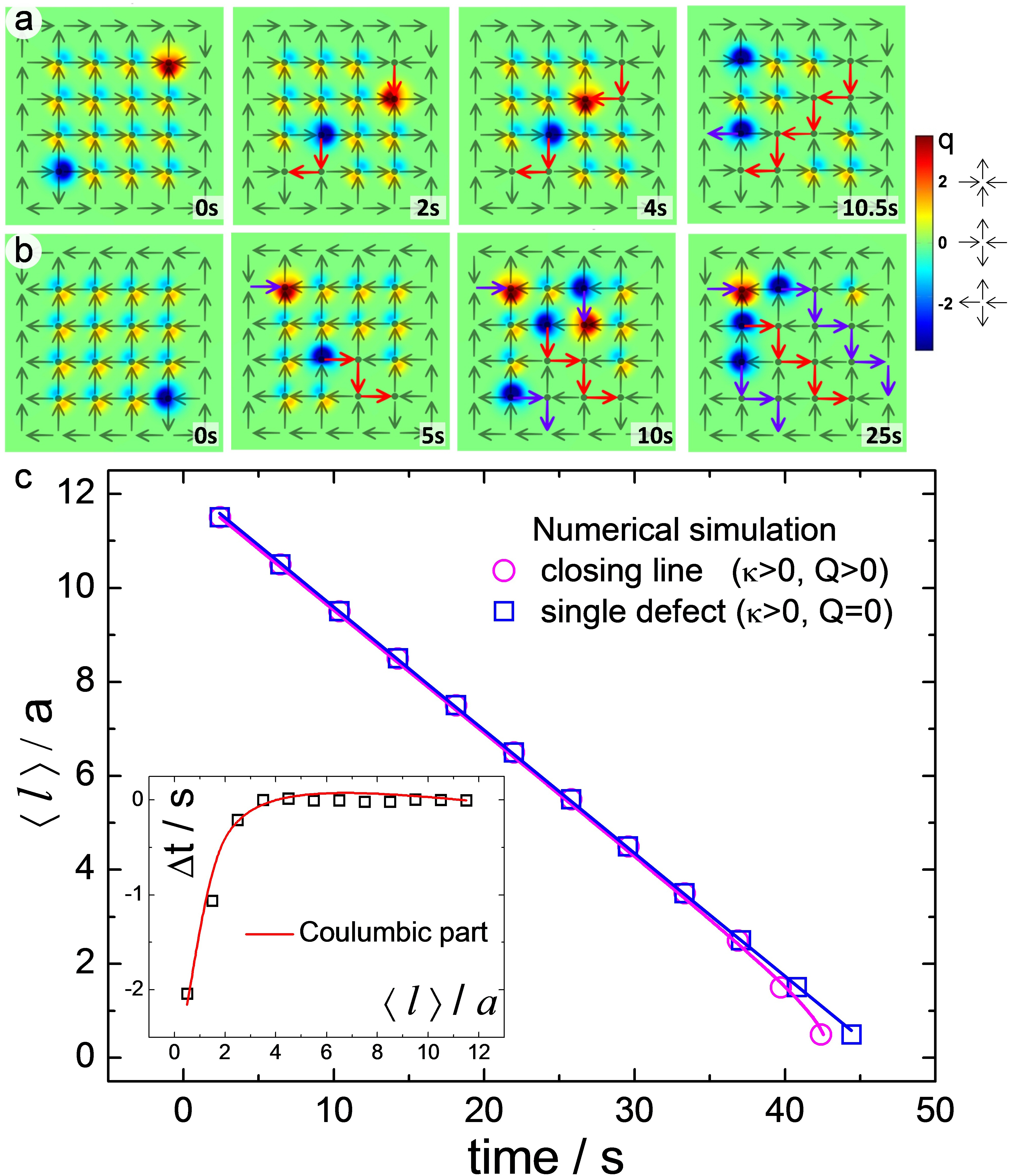}
\caption{(color online) (a,b)
Experimental vertex charges for a closing line (a) and for a single $q=-2$ propagating defect (b) in the biased state.
Gray (black) [Red (blue)] arrows are spins
flipped by
the motion of the original (spontaneously emerged) defects.
Corresponding movies (VideoS1,VideoS2) are in~\cite{EPAPS}.
(c) Numerical simulation of case (a) (with one end fixed) and (b)
showing the evolution of the line length $\langle l \rangle$
for an applied field $B_z=18.8 {\rm mT}$.
Continuous lines are fits from Eq.(1).
Bottom inset: difference between the two curves in the main panel (empty squares) 
versus line length plotted with Eq.(1) with $\kappa=0$ (continuous line).}
\label{figure3}
\end{center}
\end{figure}
analysis,  we complement the experimental measurements with Brownian dynamics simulation, following the scheme 
described in~\cite{EPAPS}.
In the simulation, we use the same experimental parameters and disorder level,
\begin{figure*}[t]
\begin{center}
\includegraphics[width=\textwidth,keepaspectratio]{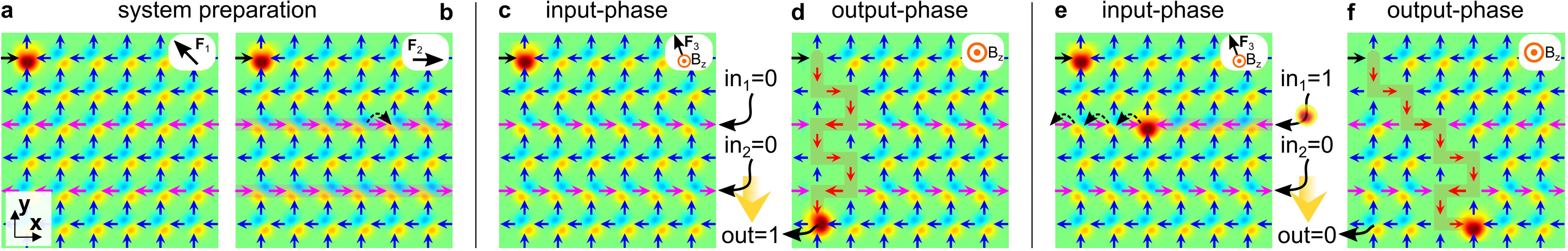}
\caption{(color online) Realization of a NOR gate via numerical simulation. (a,b) Images showing the
system preparation: (a) a force $F_1$ applied along the diagonal biases the system,
except for one fixed spin,
(b) a smaller force $F_2$ shifts two rows of particles (magenta) with high magnetic susceptibilities.
(c,d) Images showing a $1$ output obtained from $(0,0)$ inputs.
In (c) an external field $B_z$ is applied perpendicular to the plane to
start defect propagation, while a small force $F_3$ along the diagonal
prevents motion of other defects from the upper left corner.
The final result shown in (d) is the $1$ output. (e,f) Images showing a $0$ output obtained from $(1,0)$ input.
In (e) an input current causes the
whole line of magenta spin to flip ($1$ input).
The $0$ output results from the changed trajectory 
of the propagating defect (f).
All the corresponding videos can be found in~\cite{EPAPS}.}
\label{figure4}
\end{center}
\end{figure*}
and find again very good agreement with the measured data, Fig.2(b).\\
We can clearly visualize the effect of the magnetic 
Coulumbic contribution by studying defect motion in the biased system, which can be prepared 
by displacing all particles towards one of the system corners with the optical tweezers.
In this state, it is possible to generate defect lines characterized
by positive or negative line tension, or single defects with
zero Coulumbic contribution that propagate along a diagonal~\footnote{It is possible
to create a single topological monopole
which propagates in the bias state purely due to line-tension
because of the finite system size.
Once a charged defect is created, an oppositely charged one
is needed in order to fulfill charge conservation.
The flipped spins simply place one of the two defects outside the border region,
such that it does not interact with the system.}.
Of these three cases, Figs.3(a,b) shows the first and the last one,
the rest is in~\cite{EPAPS}.
The first case is shown in
Fig.3(a), where two $q=\pm 2$ defects approach when an external field $B_z=25.7 {\rm mT}$
is applied, leaving a series of $S_{III}$ vertices behind.
This situation is similar to the defect motion in the GS, with
attractive line tension and Coulumbic interaction.
We also calculate the interaction potential
$V(l)$ (data not shown), obtaining an almost identical
plot as the inset in Fig.2(c).
In contrast, in Fig.3(b) a single $q=-2$ defect propagates along the lattice only
due to line tension, since the absence of other charges sets the Coulumbic term in Eq.(1) to zero.
In the bias state we find that the defect dynamics are much slower
than in the GS, and usually in the experiments the particles stop 
propagating due to disorder, Fig.1(f).  We thus cannot directly measure
the small Coulombic contribution
in this state, however we can resolve it via  
simulation by using the same experimental conditions as in Fig.3(a,b) 
and a much larger, disorder free system. The result of these 
simulations are shown in
Fig.3(c), where we
compare the motion of single ($Q_M=0$) and double
defects ($Q_M>0$).
Both have the
same line-tension contribution and therefore move at an
identical speed for large distances. However the closing
line speeds up when the two defects are approaching at
the end of the process due to sole Coulombic interaction.
This time difference
is shown in the inset of Fig.3(c),
and can be well fitted by Eq.(1) with $\kappa=0$
(continuous line),
resulting in a similar value for the topological Coulumb charge as in the GS.\\
A major driving interest in studying defect dynamics in ASI lays
on the possibility of realizing dissipation-free "magnetronic" circuitry~\cite{Gib11,Nis13}.
We demonstrate that the colloidal spin ice system can be 
used to perform logic operations based on the motion of topological monopoles defects.
Figure 4 shows the realization, via numerical simulation, of a "NOR" gate, which is a
functionally complete port capable of generating all logical functions~\cite{Bar91}.
The gate is completely resettable, since it requires only external 
fields or gradients to work, and not individual manipulation via laser tweezers.
It is realized in a biased system, which could be formed and reset
by an external magnetic force $\bm{F} \sim (\bm{B} \cdot \nabla \bm{B})$ 
applied along one diagonal direction, $\bm{F}_1= F_1 (\hat{\bm{y}}-\hat{\bm{x}})$.
In the preparation step (Fig.4(a)) the system is biased by a 
force $F_1=2.8 {\rm pN}$ which displaces all
particles except for a pinned one which represents a fixed spin, (top left corner in Fig.4a).
We use a second type of paramagnetic colloids with a higher magnetic 
susceptibility, $\chi_2$ and ratio $\chi_2/\chi_1=1.15$,
a prerequisite which forced us to restrict the realization to the sole numerical scheme.
These particles are placed along two parallel rows spaced by two lattice
constant (magenta arrows in Fig.4).
In the second preparation step (Fig.~\ref{figure4}(b)), these particles 
are selectively manipulated by a small in-plane force
$\bm{F}_2= F_2 \hat{\bm{x}}$, $F_2=1.6 {\rm pN}$ while all other 
particles ($\chi_1$) remain at rest.
The two rows represent the inputs of the logic gate: a $0$ ($1$) is 
associated with a shifted (un-shifted) row.
After preparation of the system, a $B_z=15 {\rm mT}$ field perpendicular to 
the plane induces the defect propagation, Figs.(c-f).
The output of the gate is measured at the bottom left corner of the sample: 
it is $1$ if there is a magnetic current, $0$ otherwise.
Figures 4(c-d) describe the situation of the input $(0,0)$ with output $1$, 
while Figs.4(e-f) have input $(1,0)$ and output $0$.
In the third step (Fig.4(c) and 4(e)) a small locking 
force $\bm{F}_3=F_3( \hat{\bm{y}}-\hat{\bm{x}})$, $F_3=0.7 {\rm pN}$
is applied to hold the defect in place while the input is prepared.
Now let's consider the case of the $(1,0)$ input: an applied magnetic
current causes the upper first magenta line to flip back
into the $x<0$ direction.
In Fig.4(f), $F_3$ is set to zero and the defect starts moving.
When it reaches the flipped input row, the defect changes its path, ending 
in a different place, thus the output is $0$.
Since only $(0,0)$ input gives a $1$ output,
our logic port behaves as a NOR gate.
A similar system could be engineered in nanoscale ASI using islands of different size or 
magnetic materials
which would give spins that
behave differently under an external field.
In this context, a recent work demonstrated
the possibility to
reorient the magnetization of
the nanoislands in ASI with
an MFM tip~\cite{Wang962}.\\
In summary, we studied the defect dynamics in an artificial colloidal spin ice in the GS and in the biased state
and directly measure their energetic contributions. Our findings also confirm former theoretical
assumptions and clearly demonstrate that
these defects behave like
bound magnetic monopoles.
We finally demonstrate a resettable functionally complete "NOR" gate.
The possibility to control topological monople defects in spin ice states may foster the realization of
novel memory and logic devices based on magnetic current~\cite{Blu12,Gib11,Gil15}.\\
We thank Andras Libal and Demian Levis for stimulating discussions.
This work was supported by the ERC StG No. 335040.
P. T. acknowledges support from Mineco (Project No. FIS2013-41144-P) and AGAUR (Project No. 2014SGR878).
\bibliography{biblio}
\end{document}